\documentclass[pra,showpacs,twocolumn]{revtex4}
\usepackage{amsfonts}
\usepackage{amsmath}
\usepackage{amssymb}
\usepackage{bm}
\usepackage{epsfig}
\usepackage{graphicx,graphics}
\usepackage{wrapfig}
\usepackage[german,english]{babel}

\begin{document}

\title{Universal quantum computer from a quantum magnet}
\author{Jianming Cai$^{1,2}$, Akimasa Miyake$^{3}$, Wolfgang
D\"{u}r$^{2}$ and Hans J. Briegel$^{1,2}$}
\affiliation{$^1$Institut f\"ur Quantenoptik und Quanteninformation
der
\"Osterreichischen Akademie der Wissenschaften, Innsbruck, Austria\\
$^2$Institut f{\"u}r Theoretische Physik, Universit{\"a}t Innsbruck,
Technikerstra{\ss }e 25, A-6020 Innsbruck, Austria\\
$^3$Perimeter Institute for Theoretical Physics, 31 Caroline St. N.,
Waterloo ON, N2L 2Y5, Canada}
\date{\today}

\begin{abstract}
We show that a local Hamiltonian of spin-$\frac{3}{2}$ particles
with only two-body nearest-neighbor Affleck-Kennedy-Lieb-Tasaki and
exchange-type interactions has an unique ground state, which can be used
to implement universal quantum computation merely with single-spin
measurements. We prove that the Hamiltonian is gapped, independent
of the system size. Our result provides a further step towards
utilizing systems with condensed matter-type interactions for
measurement-based quantum computation.
\end{abstract}


\maketitle

\section{Introduction}

Quantum computers are believed to be more powerful than their classical
counterpart, resulting in tremendous efforts to implement quantum
computation with different physical systems. The model of one-way
quantum computer \cite{Rau01,Bri09} has opened a novel approach towards
the possible experimental realization of quantum computation. In the
one-way model, quantum computation starts from preparing certain universal
resource states, namely the cluster states \cite{Bri01}, and is achieved by merely
performing single-qubit measurements on these states. The principal
task turns out to be the preparation of these highly entangled
resource states. One straightforward way is to apply entangling
gates to couple a lattice of qubits \cite{Blo03}. It is however
clearly appealing if there exists a universal resource as the unique
ground state of a naturally occurring gapped two-body local
Hamiltonian, for the advantage of flexible state preparation as well
as the stability against the local perturbations.

Recall that the cluster states cannot occur as a non-degenerate
ground state of any two-local spin-$\frac{1}{2}$ Hamiltonian
\cite{Nes08}, and have singular entanglement features, e.g.
vanishing two-point correlation functions \cite{He06}. Perturbative
Hamiltonians have been proposed with (encoded) graph states as the
approximate ground states \cite{Ba06}, which however require a
highly precise control over system parameters and the spectral gap
would get significantly smaller according to the order of the
perturbation. Recently, a gapped two-body Hamiltonian of six-level
particles \cite{Chen08} has been constructed with the so called
tri-cluster state as its unique ground state, which is universal for
measurement-based quantum computation. Here, we construct a new type
of gapped local Hamiltonian, with the constitutional two-body
Affleck-Kennedy-Lieb-Tasaki (AKLT) \cite{AKLT88} and exchange-type
interactions acting on four-level particles. The simplicity of the
Hamiltonian might allow one to identify possible realizations in
models of condensed matter systems or in quantum optical set-ups.
The present approach differs from the previously used methods to
construct a parent Hamiltonian for projected entangled pair states
by finding the local support subspace \cite{Chen08,San09}, but can
be utilized to construct a family of local Hamiltonians with the
same properties.

In this paper, motivated by the usage of the gapped ground state of
the 1D AKLT spin-1 chain as a universal quantum wire \cite{Aki08}
(cf.\cite{Gr07prl,Gr07pra}), we start from 1D {\it quasi-chain} of
spin $\frac{3}{2}$ particles as a building block. From there we
construct a full 2D resource that allows a deterministic decoupling
of the 1D quantum wire structure (in the similar way as the cluster
state). After proving that the 1D quasi-chain is gapped with an
unique ground state,  we construct a gapped two-body Hamiltonian on
an octagonal 2D lattice, which is transitionally invariant, and
consists of only nearest neighbor AKLT- together with exchange-type
interactions. We demonstrate how to implement universal quantum computation by
simply making single-spin measurements on the individual four-level
particles of its unique ground state. This Hamiltonian provides the
example that can be utilized as a complete measurement-based
ground-code quantum computer without the demand of dynamical
coupling \cite{Aki08}. Extensions of our approach to the other
geometric configurations, e.g. 2D lattice, or other Hamiltonians are
also possible.

\begin{figure}[b]
\epsfig{file=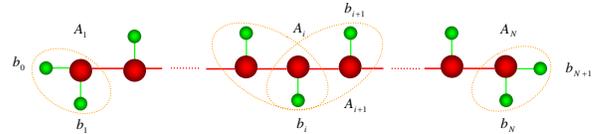,width=8cm} \caption{Configuration of a 1D
AKLT quasi-chain, which consists of spin-$\frac{3}{2} $ (red) and
spin-$\frac{1}{2}$ (green) particles. The Hamiltonian is regrouped
into blocks, each of which is marked by the dotted circle.}
\label{Qcha}
\end{figure}

\section{1D AKLT quasi-chain} We first consider the 1D AKLT model
defined on the quasi-chain as in Fig.~{\ref{Qcha}} and show that the model is gapped and has a unique ground state.
The quasi-chain consists
of spin-$\frac{3}{2}$ (A) and spin-$ \frac{1}{2}$ particles (b)
coupled with nearest-neighbor two-body interactions
\begin{equation}
H=J[\sum\limits_{i=1}^{N-1}\mathrm{P}^{3}_{A_{i},A_{i+1}}+\sum\limits_{i=1}^{N}\mathrm{P}^{2}_{A_{i},b_{i}}+\mathrm{P}^{2}_{A_{1},b_{0}}+\mathrm{P}^{2}_{A_{N},b_{N+1}}
] \label{ChainHamil}
\end{equation}
where $\mathrm{P}^{S}_{m,n}$ represents the projector onto the
spin-S irreducible representation of the total spin for particles
$m$ and $n$ (cf. \cite{AKLT88, Xu08}). As the spin per particle equals to half of the local
coordination number (i.e. the number of the bonds from a particle),
the 1D AKLT quasi-chain in Eq.~(\ref{ChainHamil}) has an unique
ground state $|\mathcal{G}\rangle $ \cite{KLT88}, which can be
obtained via a projector which maps the symmetric part of three
spin-$\frac{1}{2}$ (virtual qubits) of maximally entangled pairs
into one spin-$\frac{3}{2}$ physical particle.

To prove that the 1D AKLT quasi-chain is gapped in the thermodynamic
limit, we first regroup the Hamiltonian in Eq.~(\ref{ChainHamil})
into blocks as $H=\sum_{i}\Pi_{i,i+1}$, see Fig~\ref{Qcha},
where
\begin{equation}
\Pi_{i,i+1}=P^3_{A_i,A_{i+1}}+\frac{1}{2} P^2_{A_i,b_i} + \frac{1}{2}P^2_{A_{i+1},b_{i+1}}
\end{equation}
is the sum of the AKLT interactions among the
particles inside the block. We consider the ground state $|\mathcal{G}\rangle$ and denote the support subspace of the
reduced density matrix corresponding to each block as $ \mathcal{S}_{i,i+1}$, and the projector onto the orthogonal subspace by $\mathcal{S}_{i,i+1}^{\bot}$. From $\mathcal{S}_{i,i+1}=\ker  \Pi_{i,i+1}$ it follows that $|\mathcal{G}\rangle $ is also the unique ground
state of the projective Hamiltonian $H_{p}=\sum_{i=0}^{N}\mathcal{S}_{i,i+1}^{\bot}$.
One can directly calculate the energy gap $\gamma$ of $\Pi_{i,i+1}$
and finds $\gamma\geq 0.3518$,
which leads to
\begin{equation}
\Pi_{i,i+1} \geq J \gamma \mathcal{S}_{i,i+1}^{\bot}
\end{equation}
We further write the block Hamiltonian from $H_{p}$ on $n+1$ units as $h_{n,i}=%
\sum_{j=i}^{i+n-1}\mathcal{S}_{j,j+1}^{\bot}$. If
$h_{n,i}^{2}\geq \varepsilon h_{n,i} $, the results in
\cite{Knabe88} imply that
\begin{equation}
H_{p}^{2}\geq \frac{n}{n-1}(\varepsilon -\frac{1}{n})H_{p}
\end{equation}
Our calculations show that $\varepsilon =0.4132$ for $n=4$. We thus
conclude that $H_{p}$ is gapped as $\varepsilon>\frac{1}{4}$, where
better bounds can in principle be obtained for higher $n$.
Therefore, the energy gap of the 1D AKLT quasi-chain is lower
bounded by $ \Delta E \geq 0.0766 J$.

\begin{figure}[t]
\epsfig{file=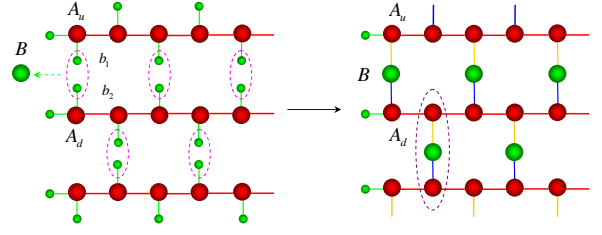,width=8cm} \caption{A 2D gapped Hamiltonian
from 1D AKLT quasi-chains. Two spin-$\frac{1}{2}$ particles ($b_{1}$
and $b_{2}$) on two neighboring chains are mapped into one
spin-$\frac{3}{2}$ particle ($B$). The nearest-neighbor interactions
consist of two-body couplings as  $E_{a} $(red), $E_{u}$ (yellow)
and $E_{d}$ (blue), $E_{b}$ (green). } \label{Conf2D}
\end{figure}

\section{2D gapped Hamiltonian by merging 1D quasi-chains} With
the above established results for the 1D AKLT quasi-chain, we will
show how to obtain a model on an octagonal lattice with two-body
interactions that is gapped and has a unique ground state. We start
from a number of independent 1D AKLT quasi-chains, and introduce the
unitary transformation $U$ which maps two spin-$\frac{1}{2}$
particles into one spin-$\frac{3}{2}$ particle by
\begin{equation}
U=\sum_{m_{1},m_{2}=\pm \frac{1}{2}}\vert
\frac{3}{2},m_{1}+2 m_{2}\rangle \langle
\frac{1}{2},m_{1}\vert\langle \frac{1}{2},m_{2}\vert ,
\end{equation}
where the labels denote $| \mathbf{S}, \mathbf{S}_z \rangle $
respectively. Notice that in principle {\em any} unitary operation
can be used at this stage, each leading to a gapped model with a
unique ground state that can be used for universal measurement-based
computation. Using such a transformation, we can merge a
number of 1D quasi-chains and get a 2D Hamiltonian as
$H_{2d}={\mathcal U} \cdot \sum_{i}H^{(i)} \cdot {\mathcal
U}^{\dagger }$ with ${\mathcal U} =\bigotimes_{\langle b_{k},b_{l}
\rangle \in E}U(b_{k},b_{l})$, where $\langle b_{k},b_{l} \rangle
\in E$ denote two neighboring spin-$\frac{1}{2}$ particles in the
same merging circle, see Fig.~\ref{Conf2D}. Thus,
\begin{equation}
H_{2d}=\sum\limits_{t=a,b,u,d}\sum\limits_{\langle m,n \rangle \in
E_{t}} \Pi^{t}_{m,n} \label{Hamil2D}
\end{equation}%
where $\Pi^{a}_{m,n}=\mathrm{P}^{3}_{m,n}$,
$\Pi^{b}_{m,n}=\mathrm{P}^{2}_{m,n}$, $\Pi^{u}_{m,n}=U(
\mathrm{P}^{2}_{A,b} \otimes \mathbb{I} ) U^{\dagger } $ and
$\Pi^{d}_{m,n}=U( \mathbb{I}\otimes \mathrm{P}^{2}_{b,A})U^{\dagger
}$, with $E_{a}$, $E_{u}$, $E_{d}$, $E_{b}$ represent different
types of couplings, see Fig.~\ref{Conf2D}. The coupling between $A$
and $B$-type particles,
\begin{eqnarray}
\Pi^{u}=\frac{1}{2}\mathbf{S}_{A_{u}}\cdot \mathbf{S}_{B}^{\prime }+
\frac{5}{8}\mathbb{I}, && \Pi^{d}=\frac{1}{2}\mathbf{S}_{B}^{\prime
\prime }\cdot \mathbf{S} _{A_{d}}+\frac{5}{8}\mathbb{I},
\end{eqnarray}
are effectively exchange-type interaction with
$\mathbf{S}^{\prime } =\mathbf{s}(-\frac{3}{2},-\frac{1}{2}) \oplus \mathbf{s}(+\frac{1}{2},+\frac{3}{2})$
and
$\mathbf{S}^{\prime \prime } =\mathbf{s}( -\frac{3}{2},+\frac{1
}{2}) \oplus \mathbf{s}(-\frac{1}{2},+\frac{3}{2})$
where $\mathbf{s}(\alpha,\beta)$ is the effective spin-$\frac{1}{2}$
operator defined on two levels $ \mathbf{S}_z = \alpha, \beta$.
It is easy to verify that $\mathbf{S}^{\prime }$ and
$\mathbf{S}^{\prime \prime }$ satisfy the commutation relations
analogous to the spin angular momentum.

As $H_{2d}$ is equivalent up to a unitary transformation to $N$ independent AKLT quasi-chains, the spectrum and the corresponding eigenvalues can be trivially obtained from the spectrum and eigenvalues of $H$. In particular, it follows that $H_{2d}$ is gapped (with the same constant energy gap
$\Delta E$ as $H$), and has the unique ground state
$ |\Psi\rangle ={\mathcal U} \cdot
\left(|\mathcal{G}\rangle \otimes \cdots \otimes
|\mathcal{G}\rangle\right). $

\section{Measurement-based quantum computation} 
\begin{figure}[b]
\epsfig{file=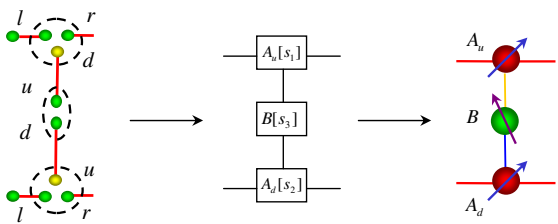,width=8cm} \caption{Computational tensor
network per single block. Entangled pairs of
virtual qubits (left) which carry logical
quantum information are mapped into physical particles (right).
The measurement on the site $B$ in different bases effectively prepares
the vertical virtual qubits (yellow) at the sites
$A_{u}$ and $A_{d}$ (d and u) into either a product
state or an entangled state, which is used to implement
decoupling 1D chains (single-qubit rotations) or two-qubit
gates, respectively. } \label{uqc}
\end{figure}

We will demonstrate how to use the resource state $|\Psi\rangle$ for
measurement-based quantum computation, in following the notation and
scheme of Refs. \cite{Gr07prl,Gr07pra,Ve04, Gr08, Cai09}. The state
$|\Psi\rangle$ can be represented as a projected entangled pair
state \cite{Ve04,Ve07}, in which a number of maximally entangled
pairs $1/\sqrt{2}(|00\rangle+|11\rangle)$ of virtual qubits are
mapped into physical particles, see Fig.~\ref{uqc} for one
computational block. The corresponding tensor matrices
for the physical site $A_{u}$ are
\begin{eqnarray}
A_{u}[+\frac{3}{2}]&=&|1\rangle _{r}\langle 0|_{l}\otimes \langle
1|_{d}\\
A_{u}[-\frac{3}{2} ]&=&|0\rangle _{r}\langle 1|_{l}\otimes
\langle 0|_{d}\\
A_{u}[+\frac{1}{2}]&=&- 1/\sqrt{3}(Z\otimes \langle
1|_{d}+|1\rangle _{r}\langle 0|_{l}\otimes \langle
0|_{d})\\
A_{u}[-\frac{1}{2}]&=&1/\sqrt{3} (Z\otimes \langle
0|_{d}-|0\rangle _{r}\langle 1|_{l}\otimes \langle 1|_{d})
\end{eqnarray}
The local tensor $A_{d}$ can be written in a similar way. The tensor
matrices for site $B$ are
\begin{eqnarray}
B[+\frac{3}{2}]&=&|0\rangle _{u}\langle
1|_{d}\\
B[- \frac{3}{2}]&=&-|1\rangle _{u}\langle
0|_{d}\\
B[+\frac{1}{2}]&=&|1\rangle _{u}\langle
1|_{d}\\
B[-\frac{1}{2}]&=&-|0\rangle _{u}\langle 0|_{d}
\end{eqnarray}
Measurement-based quantum computation on such a resource state can
be understood as follows (we refer the reader to
\cite{Gr07prl,Gr07pra,Cai09,Gr08,Ve04} for details): The logical
information in carried by the virtual qubit, and the measurement in
a certain basis on the physical particle will induce either unitary or
readout operators on the virtual qubits according to the above
tensor matrices.

Each logical virtual qubit can be initialized to $|0\rangle$ or
$|1\rangle$ by measuring the left end spin-$\frac{1}{2}$ particle in
the $\hat{z}$ basis $\{|\pm \frac{1}{2}\rangle\}$. Quantum
computation then proceeds gradually from left to right by measuring
each computational block ($A_{u}$, $A_{d}$ and $B$), see
Figs.~(\ref{Conf2D},\ref{uqc}).

Before performing computational measurements on each block, we need
to introduce a \emph{pre-normalization step} as follows to equalize
the different coefficients (1 \emph{vs.} $1/\sqrt{3}$) in the tensor
matrices $A_{u}$ ($A_{d}$). We first apply the local filter
operation $\{ L,\bar{L}\} $ on sites $A_{u}$ and $A_{d}$ with
\begin{eqnarray}
L&=&\mbox{diag}\{1/\sqrt{3},1,1,1/\sqrt{3}\}\\
\bar{L}&=&\mbox{diag}\{\sqrt{2/3},0,0,\sqrt{2/3}\}
\end{eqnarray}
An outcome
$\bar{L}$ corresponds to an unsuccessful filter attempt and leads
--after suitable measurements on sites $B,A_u,A_d$-- to the
transport of quantum information to the next block up to a local
Pauli operation $X$ or $ZX$. The procedure can hence be repeated
until it succeeds (i.e. the outcome $L$ occurs). One finds that $l\sim
O(\log\frac{1}{\epsilon })$ trials lead to an overall success
probability of $p_{s}=1-(1/3)^{l}\geq 1-\epsilon $, i.e. the process
is efficient. The required measurements are given by
\begin{eqnarray}
\{|\beta
\rangle \}=1/\sqrt{2}\{|\mathbf{\mu }_{0}\rangle\pm|\mathbf{\nu
}_{0}\rangle, |\mathbf{\mu }_{1}\rangle\pm|\mathbf{\nu
}_{1}\rangle\}
\end{eqnarray}
for site $B$, where
\begin{eqnarray}
 |\mathbf{\mu }_{s}\rangle &=& 1/\sqrt{2}(|-\frac{3}{2}\rangle+(-1)^{s}|+\frac{1}{2}\rangle )\\
|\mathbf{\nu }_{s}\rangle &=& 1/\sqrt{2}(|-\frac{1}{2}\rangle +(-1)^{s}|+\frac{3}{2}\rangle )
\end{eqnarray}
and
\begin{equation}
\{|\alpha \rangle \}=\{|+\frac{3}{2}\rangle +|-\frac{3}{2}\rangle ,|+\frac{3%
}{2}\rangle -|-\frac{3}{2}\rangle ,|+\frac{1}{2}\rangle ,|-\frac{1}{2}%
\rangle \}
\end{equation}
for sites $A_{u}$ and $A_{d}$.

In the following, we will assume that the filter operation was
successful (i.e. the result $L$ was obtained). We show that a
suitable choice of the measurement in $B$ allows one to either
decouple 1D chains and obtain single-qubit operations or read-out,
or alternatively to couple two chains directly and obtain a
two-qubit gate. Two chains are decoupled by a measurement of site
$B$ in the $\hat{z}$ basis $\{|\pm \frac{1}{2}\rangle ,|\pm
\frac{3}{2}\rangle \}$, leading to vertical virtual qubits $d$ and
$u$ corresponding to sites $A_u, A_d$ respectively (see
Fig.~\ref{uqc}) to be prepared into either $|0\rangle $ or $
|1\rangle $. For example, if the vertical virtual qubit is
$|0\rangle $, the effective tensor matrices for $A_{u}$ is given by
\begin{eqnarray*}
A[+\frac{3}{2}]&=&0,
A[+\frac{1}{2}]=-\vert 1\rangle _{r}\langle 0\vert
_{l}\\A[-\frac{1}{2}]&=&Z, A[-\frac{3}{2}]=\vert 0\rangle _{r}\langle 1\vert
_{l}
\end{eqnarray*}
For other outcomes and the tensor $A_{d}$,
the reduced tensor matrices are equivalent up to a local basis
change. Without loss of generality, we use the effective 1D tensor
matrices $A$ to show how to implement arbitrary single-qubit
rotations following a similar protocol as in \cite{Aki08} (note that
the effective tensor matrices are not equivalent to the 1D AKLT
spin-1 chain), and read out logical quantum information.

\subsection{Readout} The readout is realized by simply measuring site $A$ in the $\hat{z}$
basis $\{|\pm \frac{1}{2}\rangle ,|\pm \frac{3}{2}\rangle \}$. Once
we get the outcome $|+\frac{1}{2}/-\frac{3}{2}\rangle $, one can
infer that the logical qubit is $|0/1\rangle $. Otherwise, the
logical qubit gets a $Z$ by-product operator and we repeat the above
procedure.

\subsection{Single-qubit gates} An arbitrary single-qubit operation can
be decomposed into three rotations around the $Z$ and $X$ axes with
three Euler angles. In order to implement a $Z$ rotation
$R_{z}(\theta)=|0\rangle \langle 0|+e^{i\theta }|1\rangle \langle
1|$, we measure site $A$ in the basis
\begin{eqnarray}
\nonumber
\{|\alpha _{z}(\theta)\rangle \}
=\{|+\frac{3}{2}\rangle,1/\sqrt{2}( e^{-i\theta
}|-\frac{3}{2}\rangle -|+\frac{1}{2}\rangle ) ,\\-1/\sqrt{2}(
e^{-i\theta }|-\frac{3}{2}\rangle
+|+\frac{1}{2}\rangle),|-\frac{1}{2}\rangle \}
\end{eqnarray}
The first outcome is not possible, while the fourth induce a by-product operator $Z$.
If however the outcome is the second or third one, we implement the
rotation $ R_{z}(\theta )$ with a Pauli by-product operator $X$ or $XZ$.
An $X$-rotation $R_{x}(\theta
)=|+\rangle \langle +|+e^{i\theta }|-\rangle \langle -|$ is implemented in a similar way. We apply
the local filter operation $\{ L,\bar{L}\} $ at the initial
pre-normalization step in the $\mathbf{S}_{x}$ basis $\{
|\frac{3}{2},\mathbf{S}_{x}=m\rangle , m=\pm \frac{3}{2},\pm
\frac{1}{2}\} $. $X$ rotations can then be realized by the same
protocol as the $Z$ rotations, only with the exchange of
$\mathbf{S}_{z}$ and $\mathbf{S}_{x}$ basis, where the measurement basis
on site $B$ at the pre-normalization and decoupling step
correspondingly changes as $\{|\beta \rangle \}\leftrightarrow
\{|\pm \frac{1}{2}\rangle ,|\pm \frac{3}{2}\rangle \}$.

\subsection{Two-qubit gate} The basic idea for the implementation of an entangling gate is to prepare the vertical virtual qubits $d,u$ corresponding to sites $A_{u}$ and $A_{d}$, into an entangled state by a suitable measurement of $B$.
In order to determine the kind of two-qubit gate that is implemented for different measurement outcomes, we first rewrite the tensor matrices $A_{u},A_d$ in the aforementioned basis of $|\mu_s\rangle , |\nu_s\rangle$.
We further define another basis for site $B$ as
\begin{eqnarray}
|\mathbf{\mu}^{\prime }_{s}\rangle
&=&1/\sqrt{2}(|-\frac{1}{2}\rangle
+i(-1)^{s}|+\frac{1}{2}\rangle\\ | \mathbf{\nu}^{\prime }_{s}\rangle
&=& 1/\sqrt{2} (|-\frac{3}{2}\rangle +i(-1)^{s}|+\frac{3}{2}\rangle)
\end{eqnarray}
The contracted tensor from $A_{u}[\mathbf{\mu}_{s},\mathbf{\nu}_{s}]$,
$A_{d}[\mathbf{\mu}_{s},\mathbf{\nu}_{s}]$ and
$B[\mathbf{\mu}^{\prime }_{s},\mathbf{\nu}^{\prime}_{s}]$, results in one of the two-qubit entangling gates
\begin{equation}
V_{m,n} =I\otimes I+i\sigma_{m}\otimes \sigma_{n}
\end{equation}
where $\sigma_{m,n}=X$ or $Y$.
Moreover, any Pauli by-product operator can propagate through the
above entangling gates as $ V_{m,n}\cdot ( \pi _{1}\otimes \pi _{2})
=( \pi _{1}^{\prime }\otimes \pi _{2}^{\prime }) \cdot V_{m,n}$,
where both $\pi_{i}$ and $\pi _{i}^{\prime}$ are Pauli operators.

The explicit procedure to implement a specific entangling gate is as
follows: we first measure site $A_{u}$ and $A_{d}$ in
the basis $\{|\mathbf{\mu}_{s}\rangle ,|\mathbf{\nu}_{s}\rangle\}$;
if the outcomes correspond to the desired entangling gate out of $
V_{m,n}$, we proceed to measure site $B$ in the basis
$\{|\mathbf{\mu}_{s}^{\prime }\rangle ,|\mathbf{\nu}_{s}^{\prime}\rangle \}$.
Otherwise if we do not obtain the desired outcomes from
the measurements on site $A_{u}$ and $A_{d}$, we
measure site $B$ in the $\hat{z}$ basis $\{|\pm \frac{1}{2}\rangle
,|\pm \frac{3}{2} \rangle \}$, which decouples the chains and leads to transport of quantum information
to the next block up to a by-product Pauli operator. In this case, we have to repeat the above procedure
in order to obtain the target two-qubit gate, but an arbitrary high success probability
is achievable efficiently as well.
The present protocol offers the flexibility to choose a two-qubit gate on demand from a set of
entangling gates.

\section{Ground-code quantum computation} If the bulk Hamiltonian
is maintained during the measurement-based computation as proposed
in the ground-code scheme \cite{Aki08}, the spectral gap appears to
provide certain protection again local noises, so that it may make
the computer more robust and easier to meet a stringent
fault-tolerant error threshold for quantum error correction.
However, potential advantages of the gap in protecting quantum
information is currently intensively being studied in the context of
the topological memory \cite{Topo}. The general question as to what
extent the passive Hamiltonian protection is helpful in
measurement-based computation, which is far from the equilibrium, is
more involved, and addressed elsewhere. Below, we describe the
complete scheme with the Hamiltonian present.

We first look at the residual Hamiltonian of 1D AKLT quasi-chain
after measuring the first $j$ particles,
\begin{equation}
H(j)=J[\sum_{i=j}^{N-1}\mathrm{P}^{3}_{A_{i},A_{i+1}}+\sum_{i=j}^{N}\mathrm{P}^{2}_{A_{i},b_{i}}+\mathrm{P}^{2}_{A_{N},b_{N+1}}
]
\end{equation}
It is gapped and twofold degenerate, which can encode one
logical qubit. One can show that the operators
\begin{equation}
\Sigma_{\sigma}=\bigotimes_{i=j}^{N}\{[i\sigma(+\frac{3}{2},-\frac{3}{2})\oplus
\sigma(-\frac{1}{2},+\frac{1}{2})]^{(A_{i})}\otimes\sigma^{(b_{i})}\}\otimes
\sigma^{(b_{N+1})}
\end{equation}
with $\sigma=X,Z$ form the representation of
$\mbox{su}(2)$, and the degenerate ground states are connected only
by these non-local operators. The computation on the 2D resource
equips similar robustness, as the Hamiltonian of the 2D model is
locally unitary equivalent to $N$ independent chains.
To utilize the gap protection, one needs to turn off the
interactions that couple the computational block (see
Fig.~\ref{uqc}) to the bulk together with those inside
the block, prior to any measurement for this block. As the Hamiltonian
$H_{2d}$ is frustration-free, this can be done in a constant time.
Also particles already measured need to stay decoupled from the
remaining bulk.

In a potential implementation of $H_{2d}$ with
trapped polar molecules in an optical lattice \cite{CoMol},
nearest-neighbor interactions can be turned off by changing the
potential depth of local wells which in turn suppresses the
tunneling rate between two neighboring wells. An alternative method
without turning off interactions is to apply fast measurements and
remove the particle from the system after the measurement, or drive
it to a dark state (which does not interact with neighboring
particles anymore).

\section{Extension to 2D lattices} Our approach can be extended
from the octagonal lattice to the 2D square lattice, for instance.
The corresponding 1D AKLT
quasi-chain consists of spin-$2$ particles, each of which is
connected with two spin-$\frac{1}{2}$ particles. In a similar way,
we can show that the energy gap is lower bounded by $\Delta E\geq J
\gamma \varepsilon_{p}=0.0418 J$ with $\gamma=0.241$ and $
\varepsilon_{p}\geq 0.1735$. We can merge a number of such 1D AKLT
quasi-chains into a 2D resource state as well, and the computational
protocol is similar.

\section{Summary} We have proposed a novel translational-invariant
gapped Hamiltonian of spin-$\frac{3}{2}$ particles with
nearest-neighbor two-body AKLT- and exchange-type interactions. Its
unique ground state is proved to be universal for measurement-based
quantum computation. The Hamiltonian inherits important properties
from the original AKLT model, while at the same time, it has
distinct features, e.g. a strictly proved energy gap. Further study
on such a Hamiltonian and its order parameter might reveal new
aspects of many-body physics regarding computational capability.

\section{Acknowledgements} J.M.C., W.D., H.J.B. thank M. Van den
Nest for helpful discussions and acknowledge the support from the FWF
(SFB-FoQuS, J.M.C. through the Lise Meitner Program) and the
European Union (QICS, SCALA, NAMEQUAM). A.M. acknowledges the support by the Government of Canada through
Industry Canada and by Ontario-MRI.

\end{document}